**Single-Atom Control of Arsenic Incorporation in Silicon for High-Yield Artificial Lattice Fabrication**

*Taylor J. Z. Stock\*, Oliver Warschkow, Procopios C. Constantinou, David R. Bowler, Steven R. Schofield, Neil J. Curson\**

T. J. Z. Stock, O. Warschkow, P. C. Constantinou, D. R. Bowler, S. R. Schofield, N. J. Curson
London Centre for Nanotechnology, University College London, 17-19 Gordon Street, London, WC1H 0AH, United Kingdom
E-mail: t.stock@ucl.ac.uk, n.curson@ucl.ac.uk

T. J. Z. Stock, N. J. Curson
Department of Electronic and Electrical Engineering, University College London, London, WC1E 7JE, United Kingdom

D. R. Bowler, S. R. Schofield
Department of Physics and Astronomy, University College London, WC1E 6BT, London, UK

Artificial lattices constructed from individual dopant atoms within a semiconductor crystal hold promise to provide novel materials with tailored electronic, magnetic, and optical properties. These custom engineered lattices are anticipated to enable new, fundamental discoveries in condensed matter physics and lead to the creation of new semiconductor technologies including analog quantum simulators and universal solid-state quantum computers. In this work, we report precise and repeatable, substitutional incorporation of single arsenic atoms into a silicon lattice. We employ a combination of scanning tunnelling microscopy hydrogen resist lithography and a detailed statistical exploration of the chemistry of arsine on the hydrogen terminated silicon (001) surface, to show that single arsenic dopants can be deterministically placed within four silicon lattice sites and incorporated with 97±2% yield. These findings bring us closer to the ultimate frontier in semiconductor technology: the deterministic assembly of atomically precise dopant and qubit arrays at arbitrarily large scales.

# 1. Introduction

Semiconductor device manufacturing is steadily approaching the ultimate frontier of miniaturization: device engineering with single-atom precision. The development of tools for deterministic and repeatable positioning of single dopant atoms in semiconductors affords tremendous opportunities, opening the door to completely new types of engineered quantum materials and devices. The fabrication of artificial solid-state dopant lattices will enable the engineering of customized electronic band structures and topological states. This will unlock new avenues in condensed matter physics, not only allowing us to probe Mott-Hubbard metal-insulator transitions,[1] understand the influence of disorder on electron localization as per the scaling theory of localization,[2] and explore the attributes of topological insulators as explained by the Su-Schrieffer-Heeger (SSH) model,[3] but also to uncover previously unseen phenomena and advance our fundamental understanding in these areas. Moreover, the ability to fabricate large-scale artificial atomic arrays of dopant atoms in silicon will provide the building blocks for transformational quantum information technologies including analog quantum simulators[4] and universal quantum computers.[5–7]

The most precise technique for the deterministic placement of individual dopant atoms in silicon is scanning tunneling microscopy (STM)-based hydrogen resist lithography (HRL).[8–12] HRL uses an STM tip to pattern an inert hydrogen monolayer on the silicon surface with single-atom precision, producing a chemically-sensitive mask to spatially confine the adsorption and reaction of gas-phase dopant precursor molecules to nearly exact surface lattice sites (< 1 nm precision). A low-temperature thermal anneal is then used to incorporate the dopant atoms into substitutional lattice sites of the surface, and subsequent epitaxial overgrowth with silicon completes the process of deterministic substitutional dopant atom patterning for atomic-scale device fabrication.[13–16] This method of dopant atom positioning in silicon offers a precision that is at least an order of magnitude greater than that of its closest competitor, single ion implantation.[17,18]

Using HRL with phosphine ($PH_3$) as the dopant precursor molecule, multiple research groups have successfully fabricated single and few dopant-atom quantum electronic devices, including single-atom, single-electron transistors,[19–21] an electron-spin based, two-qubit quantum logic gate,[22] and one and two-dimensional artificial lattices.[23,24] Despite these achievements, there are fundamental limitations to the precision and scalability of the deterministic placement of phosphorus dopants in silicon by HRL. These limitations are inherent to the interaction chemistry of phosphine with the Si(001) surface. Consequently, the only devices fabricated, to date, utilizing single-dopant precision for the active components are the single-atom transistors,[19–21] while all multi-component devices have instead used non-uniform small clusters of phosphorus atoms.[15,22,23]

The placement of a single phosphorus atom from a phosphine molecule using HRL fabrication requires the removal of a minimum of six hydrogen atoms from the hydrogen-terminated Si(001) surface, exposing three adjacent pairs of silicon lattice sites. This six-site sized adsorption window, in an otherwise hydrogen terminated silicon surface, is the smallest sized adsorption window that allows the covalent adsorption of the phosphine molecule, and its subsequent full dissociation to leave one phosphorus atom and three hydrogen atoms individually bonded to the surface.[19] Full dissociation of phosphine is a prerequisite for phosphorus atom incorporation, however it has been shown that the success rate of single phosphorus atom surface incorporation resulting from the thermal annealing of phosphine adsorbed in idealized 6-site windows is less than 70%.[25,26] This low yield presents a serious limitation on scaling up to large numbers of deterministically positioned dopant atoms, and while there are ongoing attempts to overcome this issue for phosphine,[27] the problem remains substantial. To highlight the magnitude of this problem, consider that the fabrication yield ($P$) for placement of single phosphorus atoms reduces exponentially with the total number of single atoms being incorporated ($n$), as $P = 0.7^n$. For example, the probability of successfully creating a 16-element single phosphorus-atom array is $0.7^{16}$, i.e., less than 0.4%.

In the present work, we demonstrate that arsenic atom placement using arsine ($AsH_3$) as a precursor overcomes this limitation. We show that for arsine, a four-site adsorption window is sufficient and optimal for high-yield single atom arsenic incorporation and results in a deterministic incorporation probability of 97±2%, using a single-stage fabrication cycle. The reduction from a six-site to a four-site adsorption window means that arsenic can be placed within 1 in 4 silicon lattice sites in an area of ~0.6 nm$^2$, representing a 33% enhancement in the placement precision for arsenic compared to phosphorus. Furthermore, we show that unique to arsine, an iterative patterning cycle can be introduced to maintain a high yield of single arsenic donor incorporation in silicon while allowing for a further reduction in the size of the adsorption window, thus increasing the placement precision. These findings provide a pathway to the manufacturing scale-up of silicon quantum materials and devices based on deterministic positioning of arbitrarily large numbers of single dopant atoms and the construction of artificial solid-state atomic lattices.

## 2. Results and Discussion
### 2.1. Fabrication of Atomic Arrays Using Arsine and a Hydrogen Resist on Silicon

Arsenic atom dopant arrays of arbitrary size can be fabricated in silicon using arsine and HRL because the arsenic atom from any single, *partially* dissociated arsine molecule can be reliably incorporated into the silicon lattice at an adsorption window site upon annealing. The adsorption window can thus be made small enough to contain only one arsine molecule, which always incorporates. This is notably different from the case of phosphine where phosphorous incorporation can only be achieved if a phosphine molecule *fully* dissociates within the adsorption window. This necessitates a minimum 6-silicon atom sized window, which allows for routine adsorption of multiple molecules and a 30% chance that all adsorbed phosphine will desorb during the incorporation anneal, leaving behind no phosphorus.

The unique and opportune features of arsine adsorption and arsenic incorporation through HRL adsorption windows on silicon are illustrated in **Figure 1**. Here we show three examples of the deterministic positioning of arrays of single arsenic atoms into Si(001): a 2×2 array (Figure 1a,c); a 7×1 chain (Figure 1d,e); and a 4×4 array (Figure 1f-i). The hydrogen-terminated Si(001) surface consists of rows of paired silicon atoms (dimers) with one hydrogen atom bonded to each silicon. These rows are visible as vertical lines in Figure 1a-c, g-i, and horizontal lines in Figure 1d,e. Silicon atoms at the surface that do not have a terminating hydrogen atom, such as those within the lithographically patterned adsorption windows, appear bright in the images due to the existence of broken, or 'dangling' bonds (DBs) whose orbital energy lies closer to the Fermi level and therefore make a stronger contribution to the STM tunnelling current than the adjacent hydrogen-terminated silicon atoms.

The sequence of images in Figure 1a-c shows the construction of a 2×2 arsenic atom array in three stages (see labels above left-hand images). (1) Hydrogen lithography: Figure 1a shows hydrogen desorption to create a 2×2 array of adsorption windows, each consisting of $n$ DBs ($2 < n < 8$) and therefore labelled $n$DB. It is notable that the lower-left corner window extends across two dimer rows, while the other three sites are all on a single dimer row. (2) Arsine dosing: Figure 1b shows the same surface after arsine gas exposure (3 Langmuir) to produce covalently attached AsH$_x$ adsorbates ($x$ = 0, 1, or 2, discussed further below). The presence of the AsH$_x$ adsorbates is revealed by the fact that all four adsorption windows have changed appearance, leaving a new prominent single-lobed protrusion at each location. The adsorption window in the lower-left corner, the largest of the four, now features two such protrusions, indicating two different AsH$_x$ adsorbates are attached within the windowed area, one of which we identify as AsH$_2$ (see more below). (3) Arsenic thermal incorporation: Figure 1c shows the substitutional incorporation of arsenic into the silicon surface via thermal annealing. This produces arsenic-silicon heterodimers (As-Si)[28] in the surface and ejected silicon ad-atoms on the surface (ad-Si), which can assume multiple thermally-stable structures,[29] including silicon ad-dimers formed via surface diffusion of ad-atoms[30,31]

Alongside the intentionally fabricated adsorption windows, there are also a variety of native surface defects present on the surface, and spurious hydrogen desorption sites. Spurious desorption sites are the result of imperfect hydrogen termination at the sample preparation stage and/or random hydrogen desorption events during surface imaging and hydrogen desorption lithography.[32] In order to clearly delineate the intentional adsorption windows from the spurious sites and other surface defects, the STM images in Figure 1a-c are duplicated to the right of the original image, allowing us to also follow the behavior of the defect sites through each stage of fabrication.

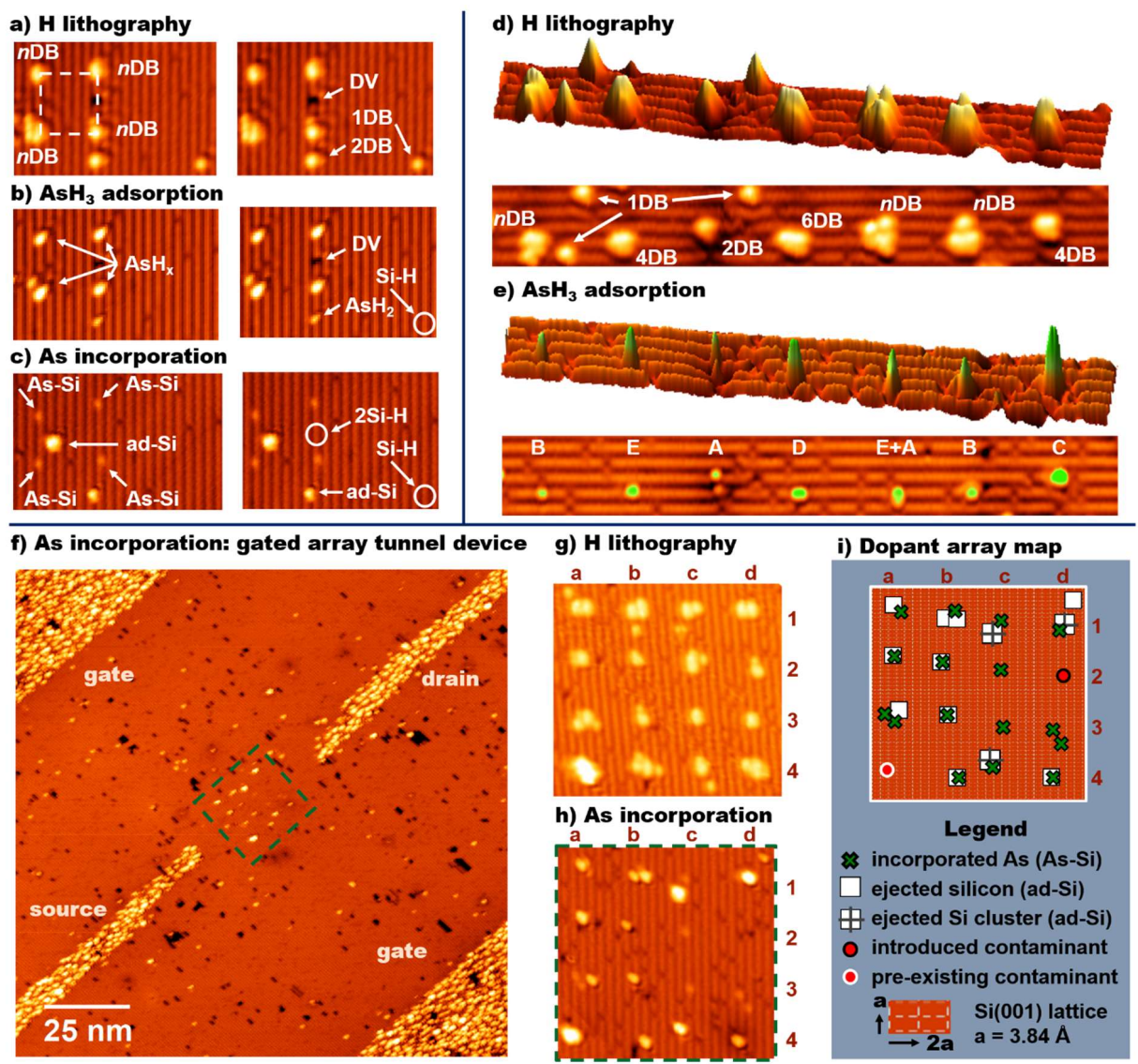

**Figure 1. Artificial lattice fabrication on Si(001) using HRL for deterministic single arsenic atom incorporation:** Three examples of single-atom arsenic arrays in silicon,

selectively imaged during the three-step fabrication process: step (1) STM lithography on hydrogen resist; (2) room temperature AsH$_3$ adsorption; (3) thermal anneal for substitutional arsenic incorporation. Panels a)-c) show all three steps for a 2×2 array, repeated images are annotated with lithographic results on the left and surface defects on the right. Labelled features: 1DB = silicon dangling bond, 2DB = dangling bond dimer, $n$DB = lithographic DB patch ($n$ = 2 to 8), DV = silicon dimer vacancy, ad-Si = silicon adatom, Si-H = hydrogen terminated silicon atom. Panels d)-e) show lithography and adsorption of a 1×7 chain. Labelled features: spurious DB (1DB), lithography sites ($n$DB), and five AsH$_x$ adsorbates (type A, B, C, D, and E features). Panel g) shows lithography and f), h) incorporation of a 4×4 atomic array within a gated source-drain tunnel device. Array elements labelled a1-through-d4 using column/row labels. Arsenic atom locations inferred from reaction products in image (h) are mapped in (i). Imagining parameters: a) -2.0 V, 40 pA; b), c) -2.0 V, 100 pA; d), e) -2.0 V, 60 pA; f), g), h) -2.0 V, 40 pA. Lithography parameters: 3.5 V, 3000 pA, 50 nm s$^{-1}$. AsH$_3$ dosing: 5×10$^{-9}$ mbar × 10 minutes. Incorporation anneal: 350 °C × 1 minute.

In the right-hand image of Figure 1a, we highlight a native silicon dimer vacancy (DV), and two spurious desorption sites: one single dangling bond (1DB) and a dangling bond pair (2DB). After arsine exposure (Figure 1b), we see that the stray 2DB site now also features a protrusion that we attribute to an AsH$_2$ fragment, while the 1DB defect has disappeared, i.e. re-terminated with hydrogen. After the thermal incorporation anneal (one minute at 350 °C), we see that the AsH$_2$ at the 2DB site is replaced by a silicon ad-atom, indirect evidence that the arsenic atom has incorporated at this site,[26,27] but is obscured by the ad-atom. Additionally, in the post-anneal image, we find that the DV has disappeared, leaving an intact dimer row. We suggest this is also due to the presence of ejected silicon, and that two diffusing silicon ad-atoms have settled in the vacancy, filling it in.

Considering the arsine dosing stage, shown in Figure 1b, we find that collectively the adsorption behaviors of both the patterned $n$DB sites and the spurious 1DB and 2DB sites are representative of the following general reaction trends: upon exposure to arsine gas 1DBs re-terminate with hydrogen, 2DBs frequently adsorb a single AsH$_2$ species, and larger $n$DB windows adsorb one or two AsH$_x$ species depending on the size of the adsorption window. We quantify these trends statistically in detail in sections 2.2 and 2.3.

Next, in Figure 1d,e, we show the hydrogen desorption and arsine adsorption stages of the fabrication of a one-dimensional seven-element chain. The seven patterned adsorption windows shown in Figure 1d are labelled $n$DB, 2DB, 4DB, or 6DB, while three spurious single-dangling bonds sites are labelled 1DB. The same area of the surface, after AsH$_3$ dosing, is shown in Figure 1e. Consistent with the adsorption trends stated above, the 1DB sites are all re-terminated with hydrogen after $AsH_3$ exposure and all seven lithographic windows feature one or two $AsH_x$ adsorbates. The varied sizes and shapes of the adsorption windows in this chain result in the formation of five characteristic $AsH_x$ adsorption feature types, which we label A, B, C, D, and E (Figure 1e). We assign the type-A and type-B features to two variants of an $AsH_2$ adsorption configuration: in the case of the type-A feature, the $AsH_2$ occupies a dimer-end position and the feature is therefore asymmetric about the dimer row, while in the type-B feature, we suggest that the $AsH_2$ is bridge bonded across a silicon dimer and the feature is thus symmetric about the dimer row. The type-C feature is an AsH+2H configuration where the AsH fragment is bound symmetrically across one dimer, while the two dissociated hydrogen atoms passivate the neighboring dimer. The type-D and type-E features are assigned to two variants of an As+3H configuration that only occur at sites large enough to accommodate the full dissociation of the molecule, see Supplementary Information **Figure S1** for more details. This varied set of adsorbate products demonstrates how the size and shape of the adsorption windows critically affects the dissociation path and final room-temperature stable adsorbate configuration. As discussed further below, a more uniform set of adsorption windows and adsorption products is desirable for device fabrication.

In Figure 1f-h we present STM images of the fabrication of a 16-element 4×4 single dopant atom array. Figure 1g shows the hydrogen lithography stage and the resulting 4×4 array of $n$DB adsorption windows. The array columns and rows are labelled a, b, c, d and 1,2,3,4, respectively, for position identification. Of these 16 $n$DB windows, six are contained within single silicon dimer rows (b2, b3, c3, c4, d2, d3), six extend across two dimer rows (a1, a2, b1, c1, c2, d1), and three across three rows (a3, b4, d4). The final remaining element (a4) is of considerably greater height than all the other lithographic array elements because it is a pre-existing point contaminant. As such, we exclude element a4 from the array statistics. There are also several spurious 1DB sites present in addition to the lithographic adsorption windows.

Figure 1h shows the same area after arsine dosing and the incorporation anneal at 350 °C. Single As-Si heterodimers can be observed in the five array sites a1, b1, c1, c2, c3; double As-Si heterodimers in the two array sites a3, d3; isolated ad-Si atoms in the five array sites a2, b2, b3, b4, d4; and ad-Si clusters in the two array sites c4, d1. The remaining two sites, a4 and d2, feature unidentified contaminants. As mentioned above, the unidentified feature at a4 was present before hydrogen lithography, while the unidentified feature at d2 appeared only after processing.

When fabricating artificial arrays using HRL, confirmation of dopant atom incorporation can be inferred from the reaction products imaged at each adsorption site following an incorporation anneal. In order to quantify the success rate of arsenic incorporation, we apply similar criteria to those used to assess phosphorus incorporation from phosphine in recent work by Ivie *et al* and Wyrick *et al*.[26,27] If an As-Si heterodimer is directly imaged at the location of the adsorption window, it is certain that arsenic incorporation has occurred at the site. If an ad-Si is imaged at the location of the adsorption window, it is assumed that arsenic incorporation has occurred at the site but is not directly visible due to the obscuring ad-Si. It is further assumed that no arsenic incorporation has occurred at sites of unidentified contaminants or hydrogen re-terminated sites. Based on this analysis, as shown in Figure 1i, we infer that 10 out of the 16 intended array sites contain single arsenic atoms. Two further sites contain one or more As atoms and the remaining 4 sites either contain more than one arsenic atom or an unidentified feature. Excluding site a4, which does not contribute to the adsorption/incorporation statistic, this tally of 10 to 12 single atoms across 15 array elements represents a compounded single-dopant-atom placement success rate of up to 80%. This result, using non-ideal sized adsorption windows, already offers a 10% plus improvement over the incorporation success rate expected for phosphorus, assuming perfect six-site adsorption windows.[25,26] However, below we show how further dramatic improvements to this single arsenic atom incorporation rate can be achieved.

Finally, Figure 1f shows a larger area image with the 4×4 array at the center. Also visible in this larger area image are four regions where we have incorporated many arsenic atoms into much larger rectangular adsorption windows near the 4×4 array. These extended and metallically doped regions would form the source, drain, and gate electrodes, as indicated, for inclusion of the 4×4 array in a field-effect device architecture. Arsenic incorporation into these extended regions is confirmed by the characteristic appearance of ejected silicon ad-dimer features within these areas. Similar device structures have been fabricated using phosphorus in silicon (for a 3×3 array of dopant clusters)[24] and we present this here as a proof-of-principle of the fabrication of such a device using only arsenic in silicon, with metallic doped arsenic sheets for electrodes and *single* arsenic dopants as the array elements.[28]

## 2.2. Adsorption and Incorporation of Arsenic at Thermally Generated Dangling Bonds:

In the following two sections, we present a statistical analysis of arsine adsorption and incorporation events at hydrogen resist windows of one to six DB in extent. We have prepared a large statistical distribution of 1DB and 2DB windows via thermal desorption by annealing hydrogen-terminated surfaces above 325 ºC.[33] An example surface is shown in **Figure 2**a. Here multiple DB sites are indicated with arrows and three examples (two 1DB and one 2DB) are marked with colored boxes and shown enlarged in highlighted panels alongside structural models below the main image. Figure 2b shows the same surface area as Figure 2a after room temperature saturation with arsine. After arsine exposure, all of the 1DB and 2DB sites either exhibit an adsorbed $AsH_2$ or have been re-terminated by hydrogen (occasional adsorption of contaminants not associated with $AsH_3$ are excluded from the statistics, an example is shown labelled "X" in Figure 2a,b).

At the 1DB adsorption windows, we observe only two outcomes: (1) re-termination by a hydrogen atom, or (2) adsorption of $AsH_2$ to form an asymmetric type-A feature. From a total count of 286 1DB sites, we find 89±2% have undergone re-termination with a hydrogen atom and 10±2% of 1DB sites have an adsorbed $AsH_2$ fragment. A full elucidation of the adsorption mechanisms for 1DB sites is outside the scope of this work, but we propose that our observations can be explained by the surface diffusion of $AsH_2$ in a physisorbed state, along the hydrogen-terminated dimer rows, following the dissociative attachment of a hydrogen atom to a 1DB site. Such a process would result in the hydrogen re-termination of a first 1DB site and the formation of a chemisorbed $AsH_2$ fragment as a type-A feature at a nearby, second 1DB site. We note that an analogous mechanism has recently been proposed for similar observations of hydrogen and non-local OH adsorption at DB sites in the dissociative attachment of $H_2O$ to partially-terminated Si(001) surfaces.[34]

At 2DB sites, we find three outcomes: (1) an asymmetric $AsH_2$ type-A feature; (2) a symmetric $AsH_2$ type-B feature; and (3) hydrogen re-termination. Following a statistical analysis of 92 2DB sites, we find 57±10% of them contain a type-A feature after saturation arsine exposure, 17±10% contain a type-B feature, and the remaining 26±5% of 2DB sites are found to re-terminate with hydrogen. The formation of the type-A feature can be understood as resulting from the dissociative attachment of a single $AsH_3$ molecule, with the $AsH_2$ fragment attaching at one end of the dimer and the hydrogen atom at the other. The formation of the type-B feature, where an $AsH_2$ fragment bonds symmetrically across a single dimer, can be explained by the surface diffusion of physisorbed $AsH_2$ fragments resulting from $AsH_3$ dissociation and hydrogen re-termination at 1DB sites as discussed above. While alternative explanations are possible, the observation of the type-B features at 2DB sites adds further support to the proposed mechanism involving diffusive physisorbed $AsH_2$ fragments. Finally, we suggest re-termination of 2DBs with hydrogen results from two sequential H-transfer reactions from two AsH$_3$ molecules to each of the dangling bonds in the 2DB structure.

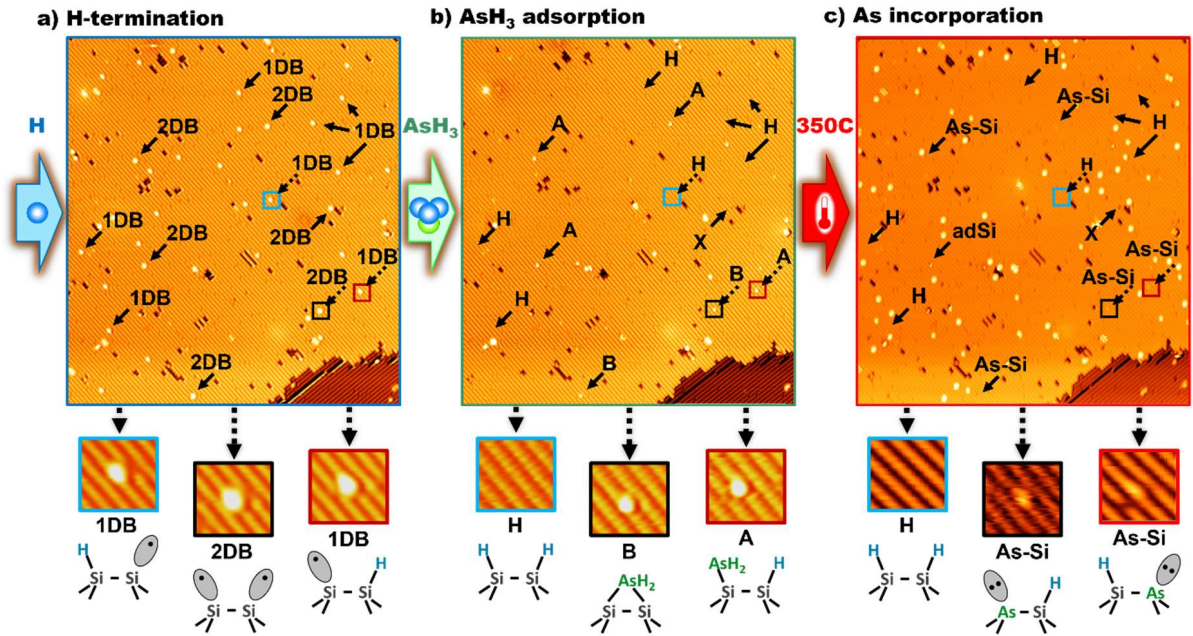

**Figure 2. Arsine adsorption and arsenic incorporation through thermally generated dangling bond sites in the hydrogen resist:** a) STM image of hydrogen terminated Si(001) containing a distribution of single (1DB) and dimer (2DB) dangling bond sites. Three highlight panels show two 1DBs identified by their asymmetric appearance, and one 2DB which is symmetric about the dimer row axis. b) The same surface after saturation with AsH$_3$. 1DB sites are re-terminated with H or adsorbed with a type-A feature (AsH$_2$ dimer-end structure). 2DBs are re-terminated or adsorbed with a type-A or type-B (AsH$_2$ dimer-bridge) feature. DBs may also infrequently adsorb non-arsenic contaminants labelled as X. c) The same surface following a 350 °C incorporation anneal. H-terminated DBs are unchanged and both type-A and type-B features are converted to As-Si heterodimers. Imaging parameters: a) b) c) -2.1 V, 20 pA. AsH$_3$ dosing: 5×10$^{-9}$ mbar × 10 minutes. Incorporation anneal: 350 °C × 2 minutes.

Figure 2c shows the same surface after a ~350 °C incorporation anneal. We find that sites adsorbed with type-A or type-B features both result in the incorporation of single arsenic atoms, as evidenced by the appearance of either an As-Si heterodimer or an ejected ad-Si atom. This is further confirmation of our assignments of the type-A and type-B features, and an indicator that arsine is an efficient dopant precursor. To quantify this efficiency, we have analyzed arsenic incorporation at a combined 156 1DB and 2DB sites, and the results are presented in **Table 1**. We find a very high probability, 94±3%, for arsenic incorporation at 2DB sites, once an AsH$_2$ fragment has adsorbed. In contrast, phosphine also adsorbs at 2DB sites as $PH_2$ in an analogous type-A feature, but shows no significant incorporation following thermal anneal,[27] and all $PH_2$ fragments are expected to desorb when annealed above 360 °C.[35] If we include the adsorption stage as well as the thermal incorporation stage, the arsenic incorporation success rate at 2DB sites drops to 70±21%, however, the failure mechanism is the re-termination of the 2DB with hydrogen during the arsine exposure, such that the 2DB site can be rewritten lithographically and the dose-anneal steps repeated. In this way, iterative lithography/dose cycles can lead to a single arsenic atom placement success rate approaching 100% at 2DB sites.

**Table 1. Reaction products of $AsH_3$ molecules at single (1DB) and dimer (2DB) dangling bond sites on hydrogen-terminated Si(001)**

| Reactive Site | Adsorbate Type | Probability | As Incorporation Probability (if $AsH_2$ adsorbed) |
|---|---|---|---|
| 1DB (single dangling bond) | H (re-termination) | 89±2% | N/A |
| 1DB | Type-A (dimer-end $AsH_2$) | 10±2% | 92±6% |
| 1DB | Type-B (dimer-bridge $AsH_2$) | 0.6±0.6% | N/A |
| 2DB (dimer dangling bond) | 2H (re-termination) | 26±5% | N/A |
| 2DB | Type-A (dimer-end $AsH_2$+H) | 57±10% | 94±3% |
| 2DB | Type-B (dimer-bridge $AsH_2$) | 17±10% | 94±3% |

The overall incorporation probability of arsenic at 1DB sites, including both the dosing and annealing stages, is 10±3%. This is too low to be useful for device fabrication while significant enough to be a potential nuisance. An obvious mitigation of this nuisance is reduction of the substrate temperature during hydrogen termination to produce a surface with a low enough density of 1DB sites (and no or few 2DBs),[36] such that spurious arsenic incorporation through ~10% of the sites poses no detriment to device fabrication. It has also recently been demonstrated that non-contact AFM can be used to re-terminate individual 1DBs, eliminating the adsorption site all-together.[37] Another possible approach is to anneal the hydrogen terminated surface allowing 1DB sites to diffuse and pair up in the more thermodynamically stable 2DBs.[38] Prior to arsine dosing, this surface can first be exposed to phosphine, effectively plugging all 2DB sites with $PH_2$ adsorbates that will eventually desorb during the incorporation anneal. By employing such mitigations, arsenic incorporation through 1DB and 2DB sites can be utilized, as needed, for incorporation of single atoms with extreme spatial precision, albeit in an iterative fabrication cycle. Alternatively, we now present further results showing how arsenic incorporation using a single fabrication cycle can be improved to 97±2% by using adsorption windows larger than 2DB.

**2.3. Adsorption and Incorporation of Arsenic at Lithographic Windows in the Hydrogen Resist:**

**Figure 3**a shows an STM image of a patterned 10×10 array of adsorption windows. These 100 lithographic sites are of various sizes, $n$DB where $n$ ranges from 2 to 10, and shapes, spanning one to three silicon dimer rows. Within this distribution, there are six 2DB, fourteen 4DB, and eight 6DB sites, all of which are confined to single-dimer rows. These 2DB, 4DB, and 6DB sites are highlighted in Figure 3a by dotted, dashed, and solid outlined circles, respectively; an example of each is shown at higher resolution with accompanying structural diagrams in Figure 3b-d. In Figure 3e, we show the same 10×10 array following a saturation dose of arsine and see that all DBs on the surface have been replaced by $AsH_x$ adsorbates or re-terminating hydrogen. Close examination of the 100 array sites reveals multiple examples of the five feature types (A, B, C, D, and E) discussed above in the seven-element chain example (Figure 1(d)), and no other $AsH_x$ adsorbate feature types. Figure 3f-h shows enlargements of the adsorbates highlighted in Figure 3b-d.

As with the thermally generated 2DB adsorption sites discussed in Figure 2, we find all 2DB sites in the 10×10 array result in type-A or type-B $AsH_2$ adsorption, or hydrogen re-termination with probabilities also in agreement with the thermal desorption experiment data. An example of type-A feature formation at a 2DB site is shown in Figure 3b,f.

At high symmetry 4DB sites, such as that highlighted in Figure 3c, we observe a preference for the formation of a type-C feature (Figure 3g). As described above, the type-C feature is an AsH fragment in a dimer-bridge configuration. It presents a protrusion that is symmetric about the dimer row, similar to the type-B feature that is a centered $AsH_2$ fragment; however, the type-C feature images both wider and brighter than the type-B feature. This structure can form at the 4DB sites when an $AsH_3$ sheds two of three H atoms to re-terminate one dimer of the 4DB, and the remaining AsH moiety bridge-bonds across the second (see schematic, Figure 3g). This type-C feature is never observed to form at 1DB or 2DB adsorption windows but is found to form at 18 of the total 24 measured 4DB adsorption windows (75%), following exposure to arsine. In addition to the type-C feature, 4DB sites are also found to occasionally adsorb single type-A or type-B $AsH_2$ + H features, or type-D or type-E As + 3H adsorbates, with probabilities of 17% and 8%, respectively. Importantly, the 4DB sites are *never* observed to contain more than one $AsH_x$ fragment, nor do any fully re-terminate with four hydrogen atoms. All 4DB sites adsorb exclusively a single AsH$_x$ fragment, i.e. from our statistics of 24 4DB sites, we find 100% adsorption probability for a *single* AsH$_x$ species, precisely what is required for scale-up to large-scale donor array fabrication.

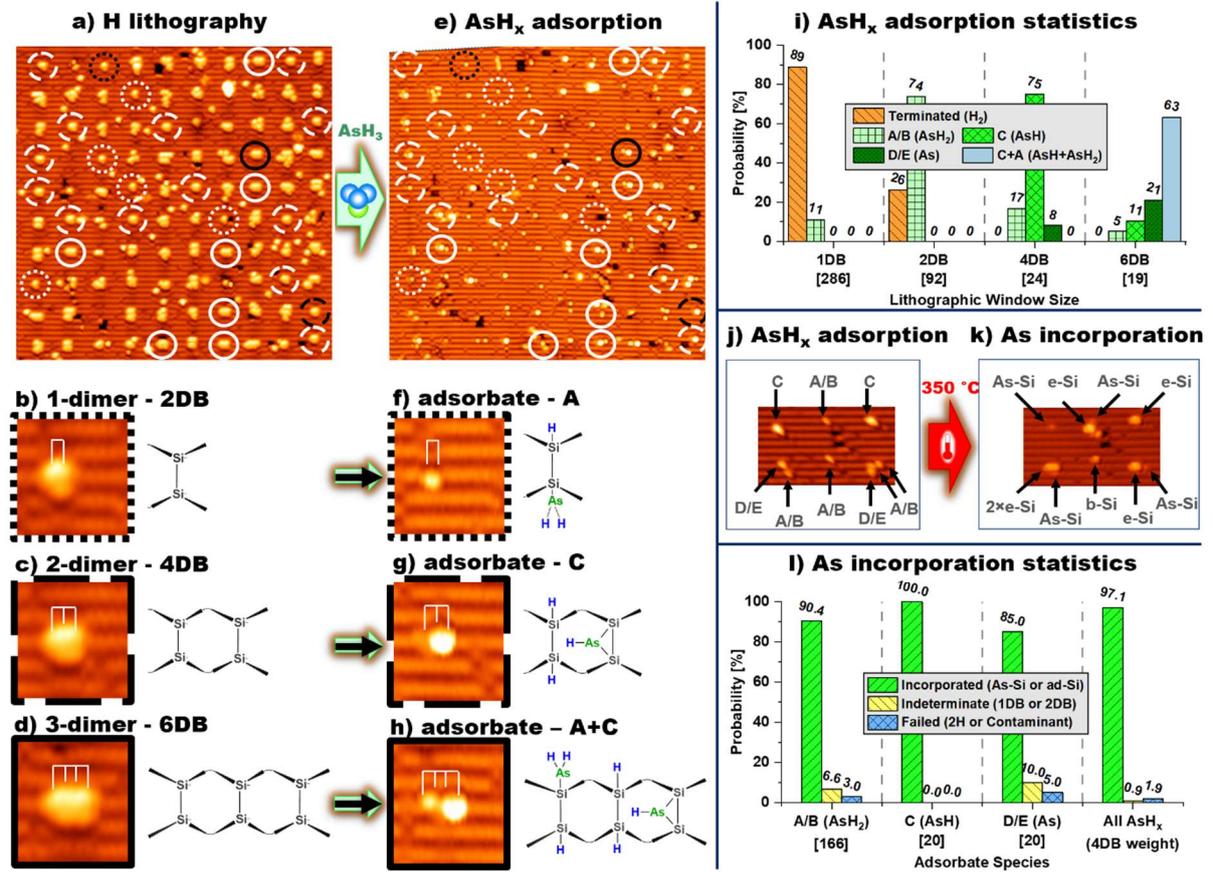

**Figure 3. Arsine adsorption and arsenic incorporation through lithographically patterned windows in a hydrogen resist:** a) STM image of a 10×10 array of *n*DB windows (*n* = 2 to 10) in a hydrogen monolayer on Si(001). Windows of size *n* = 2, 4, and 6 are highlighted by dotted, dashed, and solid bordered circles, respectively. Windows outlined in black circles are displayed in detail in panels b) c) and d) alongside corresponding structural diagrams. e) Image of the patterned area from a) following saturation AsH$_3$ exposure. All *n*DB sites are replaced with AsH$_x$ adsorbates (*x* = 0, 1, or 2) or are re-terminated. f), g), and h) show the same three array elements from d), c), and d) now filled with the adsorbates shown structurally in the accompanying diagrams. i) Adsorption statistics of the thermal 1 and 2 DB sites from section 2.2 combined with results from the 2, 4, and 6 DB lithographic sites, showing distributions of feature types A-E from Figure 1e. j) and k) show a six-element array in which all AsH$_x$ adsorbates incorporate, as confirmed by presence of As-Si, ad-Si (either dimer-bridge "b-Si" or end-bridge "e-Si") at the site of adsorption. l) Collected incorporation statistics of AsH$_x$ types A-E following 350 °C incorporation anneal. Numbers in square

brackets below histogram columns indicate total counts per species. Imagining parameters: a)-h) -2.0 V, 60 pA, j-k) -2.0 V, 100 pA. Lithography parameters: 3.5 V, 3000 pA, 50 nm s$^{-1}$. AsH$_3$ dosing: 5×10$^{-9}$ mbar × 10 minutes. Incorporation anneal: 350 °C × 1 minute.

At the locations of high symmetry 6DB adsorption windows (three contiguous 2DB sites as in Figure 3d), the preferred adsorbate structure is a pair of adsorbed AsH$_x$ fragments: a type-A feature and a type-C feature, located on neighboring dimers, as shown in Figure 3h. This AsH$_3$ adsorption structure is detected at 12 out of 19 exposed 6DB windows (63%). The remaining 37% of the 6DBs exhibit the adsorption of lone type-D or type-E features (21%), type-C features (11%) or, type-A or type-B features (5%). As with the 4DB, 6DB windows are always observed with at least one AsH$_x$ adsorbate after saturation arsine exposure.

The above adsorption behavior of AsH$_x$ fragments at 1DB, 2DB, 4DB, and 6DB adsorption windows is summarized in Figure 3i. Here, data from lithographically prepared sites are presented together with those from thermally generated 1 and 2 DB sites. The resulting adsorption probability histogram demonstrates a result of critical importance for single-atom arsenic patterning. We find that 4DB lithographic windows are the optimal size for reliable single arsine molecule adsorption, guaranteeing the adsorption of exactly one AsH$_x$ fragment. 1DB and 2DB windows are too small and routinely suffer complete re-termination. This excludes use of these sites for single-atom patterning in a single process-step sequence. In contrast, 6DB windows are sufficiently large that they preferentially adsorb two AsH$_x$ fragments, rendering them incompatible with single-atom patterning (although nevertheless useful for the type of donor cluster fabrication that has been utilized in phosphorus in silicon devices such as two-qubit gates).[22,39] The 4DB window does not completely re-terminate nor does it adsorb more than one AsH$_x$ fragment. The 4DB is therefore the perfect sized lithographic window for reliable single arsine molecule adsorption.

We next turn our attention to the statistics of arsenic atom incorporation from the various AsH$_x$ adsorbates. Figure 3j shows an example where nine AsH$_x$ fragments have been adsorbed at the locations of six lithographic adsorption windows. These fragments are isolated or clustered and are identified as five type-A or type-B, two type-D or type-E features, and two type-C. Figure 3k shows the same surface following a ~350 °C incorporation anneal. All nine adsorbates are replaced by visible As-Si, or on-site ad-Si, confirming arsenic incorporation at each site. Including the examples found in Figure 3jk, we have assessed a combined total of 206 instances of isolated type A, B, C, D, and E features after thermal annealing to 350°C (see Supporting Information), applying the same incorporation assessment same criteria discussed earlier and previously applied to phosphorus.[26,27] The results of this analysis are summarized in the histogram in Figure 3l. We find that all five AsH$_x$ adsorbate feature types (A, B, C, D, and E) provide single incorporated arsenic atoms with near unity yield. This is a remarkable result with significant implications for atomic-scale donor device scale-up.

The last panel of the histogram in Figure 3l shows the combined arsenic incorporation statistics for all five feature types where each species is weighted by its probability of adsorption at a 4DB site. From this result we expect that when any 4DB window in a hydrogen resist on Si(001) is exposed to a saturating dose of arsine and annealed to 350 °C the result will be the incorporation of a single arsenic atom with a yield of 97±2%. This is an exceptionally high incorporation yield. Furthermore, while the uncertainty of this value results from the occurrence of dangling bonds at the adsorption sites following the incorporation anneal, the 1% incorporation failure results exclusively from the re-termination of the adsorption site with hydrogen, and not from unintentional incorporation of more than one arsenic atom. This result represents a pathway to atomically precise patterning of arbitrarily large numbers of single arsenic atoms in a silicon lattice.

The preceding demonstration that arsine can overcome the shortcomings of phosphine with respect to single atom yield (~70% with 1-in-6 site precision for PH$_3$) raises an important question. Why does arsine outperform phosphine in this manner? It is known that when a PH$_2$ fragment, adsorbed at a 2DB site, is annealed to ~360 °C, it will recombine with hydrogen and desorb as PH$_3$.[35] In contrast, we have shown here that an AsH$_2$ fragment will not desorb when annealed but instead is incorporated into the Si(001) surface. We propose the following explanation of this fundamental difference between the behavior of phosphine and arsine. For any adsorbed AsH$_x$ fragment to produce an incorporated arsenic atom, it must first dissociate all its remaining hydrogen atoms to the surface; this requires available reactive surface sites. We suggest that a 350-380 °C anneal is insufficient to excite recombinative desorption of arsine but is sufficient to produce additional dangling bond sites immediately at the location of an AsH$_x$ adsorbate. At this temperature the hydrogen resist atoms become mobile,[38] and diffusion or desorption of hydrogen from the resist can deliver additional reactive silicon sites (i.e. DBs) allowing an AsH$_x$ adsorbate to completely dissociate. That is, arsine desorption has a higher activation barrier than that of DB diffusion or H$_2$ desorption, whereas the reverse is true for PH$_3$.

Due to this fortuitous surface chemistry, arsenic-in-silicon is found to be the first material system offering full control of single dopant atom incorporation. The technical difficulty of this feat is reflected in the fact that the 2×2 array of Figure 1 is the first ever atomic precision array built from single dopant atoms in silicon. Our comprehensive experiments not only demonstrate that scaling such arrays to large numbers is now possible, but our results also provide a guide to the surface chemistries that should be sought in other candidate combinations of precursors, resists, and substrates that are capable of being similarly controlled. Using this guidance, additional material systems may be added to an expanding technology exceeding the precision of all other semiconductor doping techniques and offering the capability to produce artificial lattices that meet the demanding requirements of designer quantum materials and solid-state quantum information devices. This marked improvement promises a substantial leap for device scale-up, opening the door for practical, large-scale assembly of atomically precise dopant arrays and revolutionizing quantum materials and devices.

## 3. Conclusion

We have demonstrated that by combining arsine with HRL it is possible to achieve controlled positioning of single arsenic atoms into the Si(001) surface lattice with a nominal spatial precision of 1-in-4 lattice sites and a single atom yield of 97±2%. This remarkable success rate can be achieved in a single lithography/dose/anneal cycle. The successful implementation of these findings for single atom precision device fabrication requires only a well-prepared (i.e. low DB density) surface, a high purity precursor gas, and a well-controlled STM lithography procedure capable of reliable removal of exactly four hydrogen atoms in a 2×2, double-dimer formation, as can be achieved through feedback-controlled lithography.[40–42] By introducing an iterative lithography/dose cycle it should be possible to achieve both incorporation through smaller lithographic windows containing only a single silicon dimer as well as single atom yields of up to 100%.

## 4. Experimental Methods

All scanning tunneling microscopy imaging was performed using a ScientaOmicron VT-STM operated at room temperature and a base pressure of $3\times10^{-11}$ mbar. STM tips were prepared by mechanical cutting of Pt-Ir (90:10) wire. Si(001) substrates were n-type, arsenic doped, with a nominal resistivity of 15 Ohm·cm. Clean silicon surfaces were prepared using a standard direct current heating recipe with a 600 °C × 8-hour outgas, followed by multiple ~1200 °C flash anneals, in a preparation chamber with a base pressure of $2\times10^{-10}$ mbar. Surfaces were hydrogen terminated by exposure to an atomic hydrogen beam, generated by thermally cracking of $H_2$ gas in a Tektra H-flux Atomic Hydrogen Source, using an $H_2$ partial pressure of 5×10$^{-7}$ mbar, and a substrate temperature of 320 °C maintained by direct current heating. All STM lithography was performed at a tip-sample bias of +4.0 V, a tunnel current setpoint of 3.5 nA, and a write speed of 50 nm s$^{-1}$. Sample temperatures were monitored using an Impact IR IGA 50-LO optical pyrometer, with an estimated measurement uncertainty of ± 20 °C. Total and partial pressures were measured using Bayard-Alpert ionization gauges and an SRS RGA100 quadrupole mass spectrometer.

**Supporting Information**

Supporting Information is available from the Wiley Online Library or from the author.

**Acknowledgements**

The project was financially supported by the Engineering and Physical Sciences Research Council (EPSRC) [grant numbers EP/M009564/1, EP/R034540/1, EP/V027700/1, and EP/W000520/1] and Innovate UK [grant number UKRI/75574]. P.C.C. was partly supported by the EPSRC Centre for Doctoral Training in Advanced Characterisation of Materials [grant number EP/L015277/1], and by the Paul Scherrer Institute.

**Conflict of Interest**

The authors declare no conflict of interest.

**Data Availability Statement**

The data supporting the findings of this study are openly available at zenodo.org.

# Supporting Information

**Single-Atom Control of Arsenic Incorporation in Silicon for High-Yield Artificial Lattice Fabrication**

*Taylor J. Z. Stock\*, Oliver Warschkow, Procopios C. Constantinou, David R. Bowler, Steven R. Schofield, Neil J. Curson\**

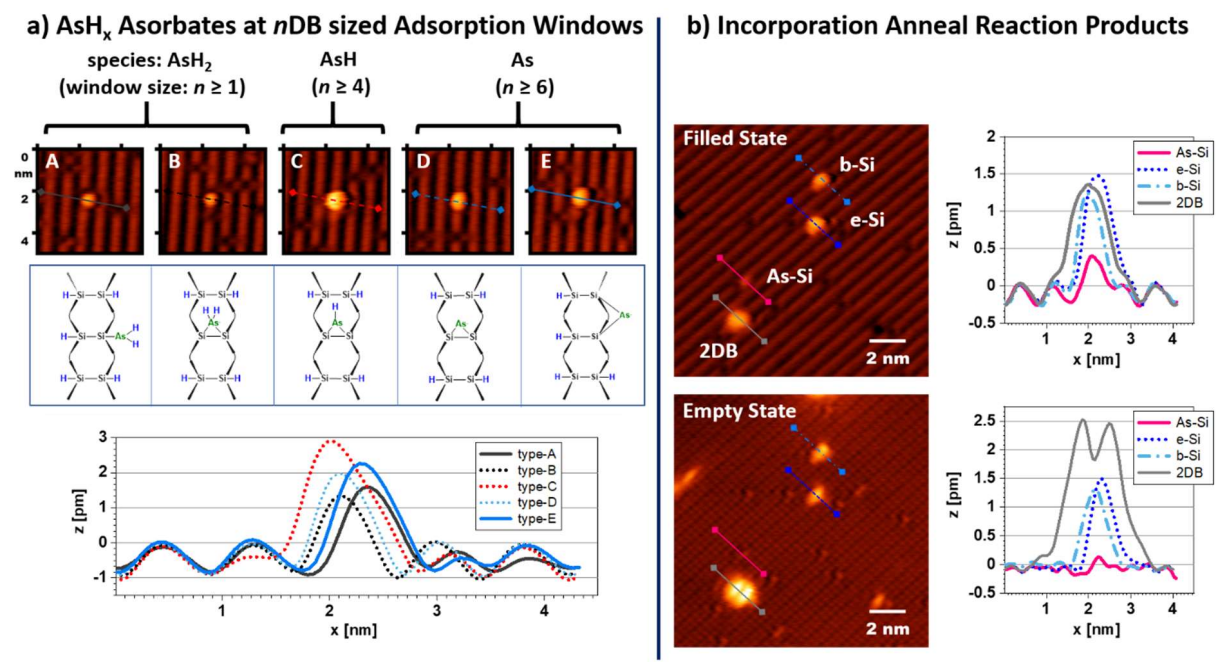

**Figure S1. AsH$_x$ Adsorption Species and Thermal Anneal Reaction Products:** When 1 to 6 DB sized hydrogen resist adsorption windows are exposed to AsH$_3$ gas, a variety of AsH$_x$ adsorbate features are detected, dependent on the window size. Panel a) shows filled state STM images and extracted line profiles of the five regularly observed adsorbate feature types, labelled A-E. These five species are divided into three categories based on the assumed degree of dissociation, corresponding to AsH$_x$, where x = 0, 1, or 2. Possible structure diagrams are provided below the corresponding images for each feature type. Species pairs A/B and D/E can only be distinguished in STM images displaying atomic resolution. When differentiation of these pairs is not possible, the full adsorbate set collapses into three species designated as A/B (AsH$_2$), C (AsH), and D/E (As)  When annealed above 350 °C, the five AsH$_x$ adsorbates can produce a variety of reaction products. Panel b) shows filled and empty state STM images and extracted line profiles of four possible products: adsorbed silicon monomers e-Si (end-bridge) and b-Si (dimer bridge), arsenic-silicon heterodimer (As-Si), and silicon dangling bonds (2DB). The above identification of adsorbed reactants and products in STM images provides a key to assessing sequential images of H:Si(001)-(2×1) surfaces that have undergone all or some of the three-step process of lithography-dose-anneal.

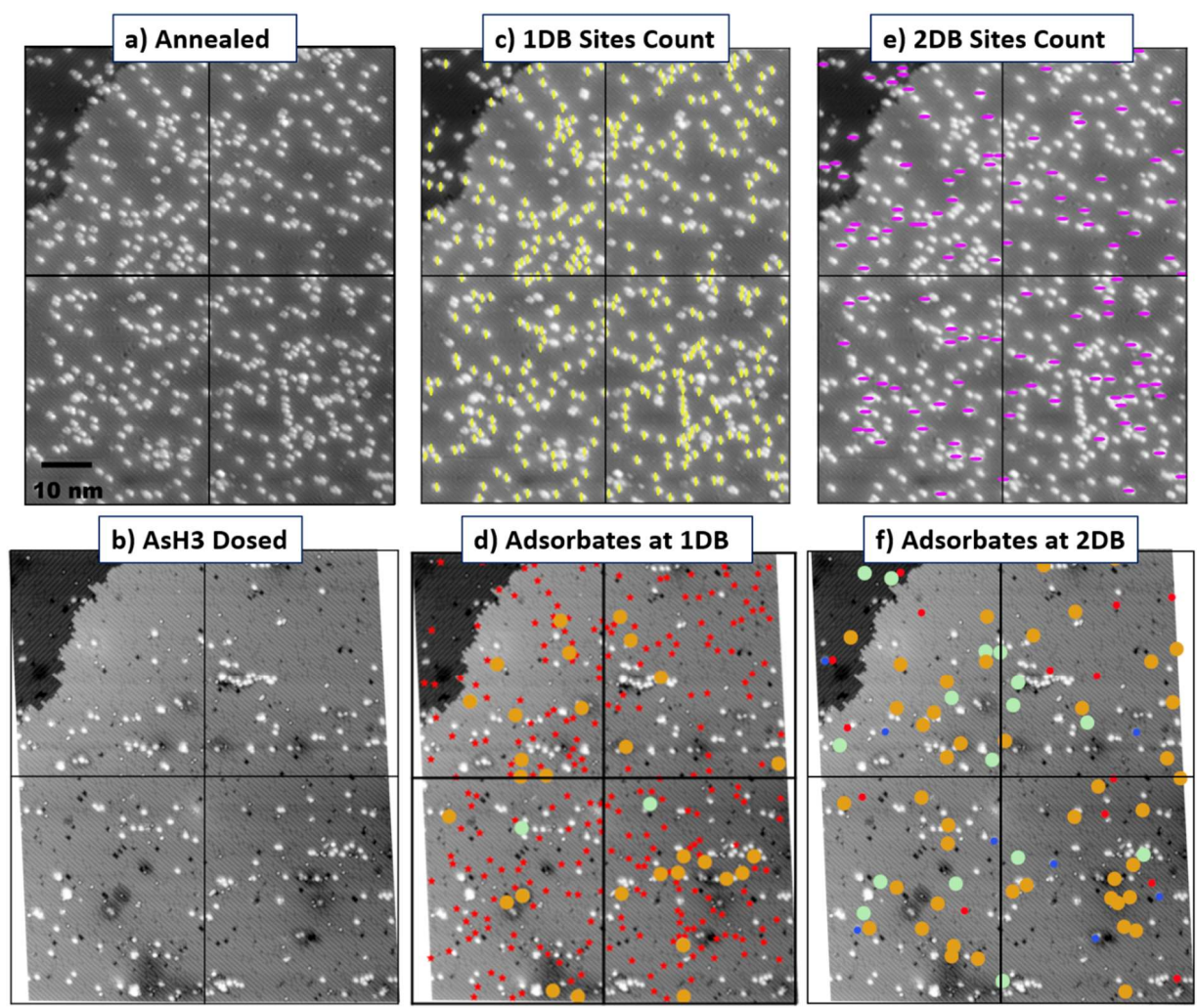

**Figure S2. AsH₃ Adsorption Statistics at 1DB and 2DB Adsorption Sites:** The same 100×80 nm area of an H:Si(001) surface is shown in panel a) following a ~375 °C anneal, and then in b) following a 3 L AsH$_3$ exposure. The anneal produces a high density of 1DB and 2DB sites, and the AsH$_3$ exposure converts these sites into AsH$_3$ related adsorbates. In panel c) all 1DB sites are marked with yellow vertical strikes, and in panel e) all 2DB sites are marked with purple horizontal strikes. In panels d) and f) the marked 1DB and 2DB sites, respectively, are followed through their conversion following AsH$_3$ adsorption. Adsorbate feature type-A are marked with a large brown circle and type-B with a large green circle. Hydrogen terminated sites are marked with a small red circle (or blue when a 2DB site is converted to a 1DB). Unexpected occasional adsorption of type-B features at 1DB sites is attributed to conversion of those 1DB sites to 2DB during the imaging process via spurious STM tip desorption (evidence of such tip desorption is found in the vertical strips of clustered adsorbates in the AsH$_3$ dosed image). The statistics of Table 1 and Figure 3i are generated by identifying and then counting the DB sites and adsorbates as illustrated here.

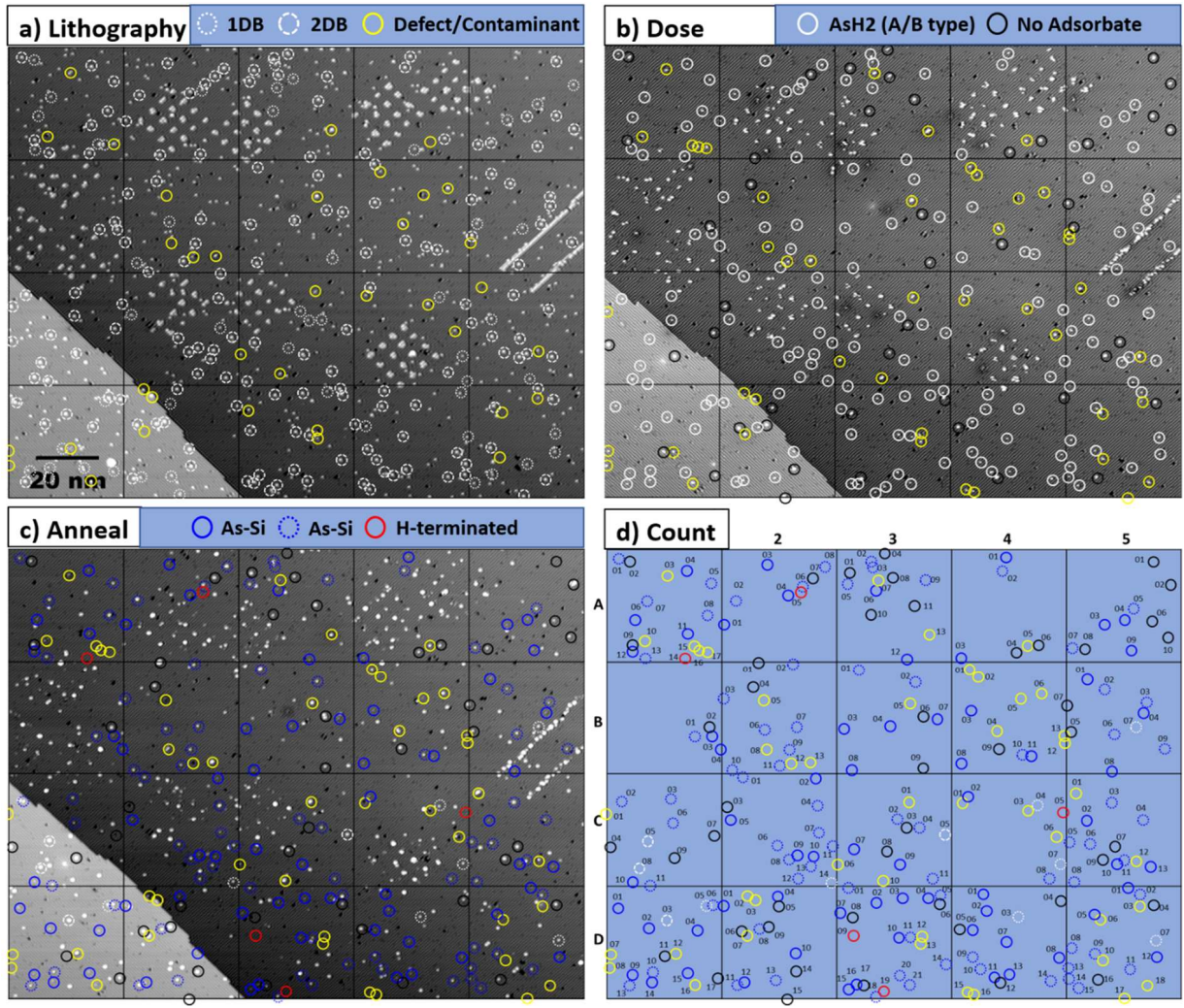

**Figure S3. Arsenic Incorporation Statistics at 1DB and 2DB Adsorption Sites:** The same 190 ×150 nm area of a H:Si(001) surface is shown following a) lithographic patterning, b) $AsH_3$ dosing (3 L), and c) thermal annealing (~350 °C) for dopant incorporation. In image (a), thermally generated 1DB and 2DB defects are marked with dotted or dashed white circles, respectively, and pre-existing defects/contaminants (including DB sites immediately next to defects/contaminants) are marked with solid yellow circles. In image (b) 1DB and 2DB sites are assessed for adsorption and marked with a solid white circle for a type A/B $AsH_2$ adsorbate, solid black circle for no adsorbate, or solid yellow circle as before (note: two A/B features immediately next to each other are counted as a defect). In image (c) the $AsH_2$ adsorbates are assessed for incorporation and marked with a solid blue circle for a As-Si heterodimer, dashed blue circle for ad-Si, dashed/dotted white circle for DBs as before, and a solid red circle for hydrogen termination. Finally, in panel d) all recorded features are counted. The statistics of Table 1 and Figure 3l are generated following this procedure.

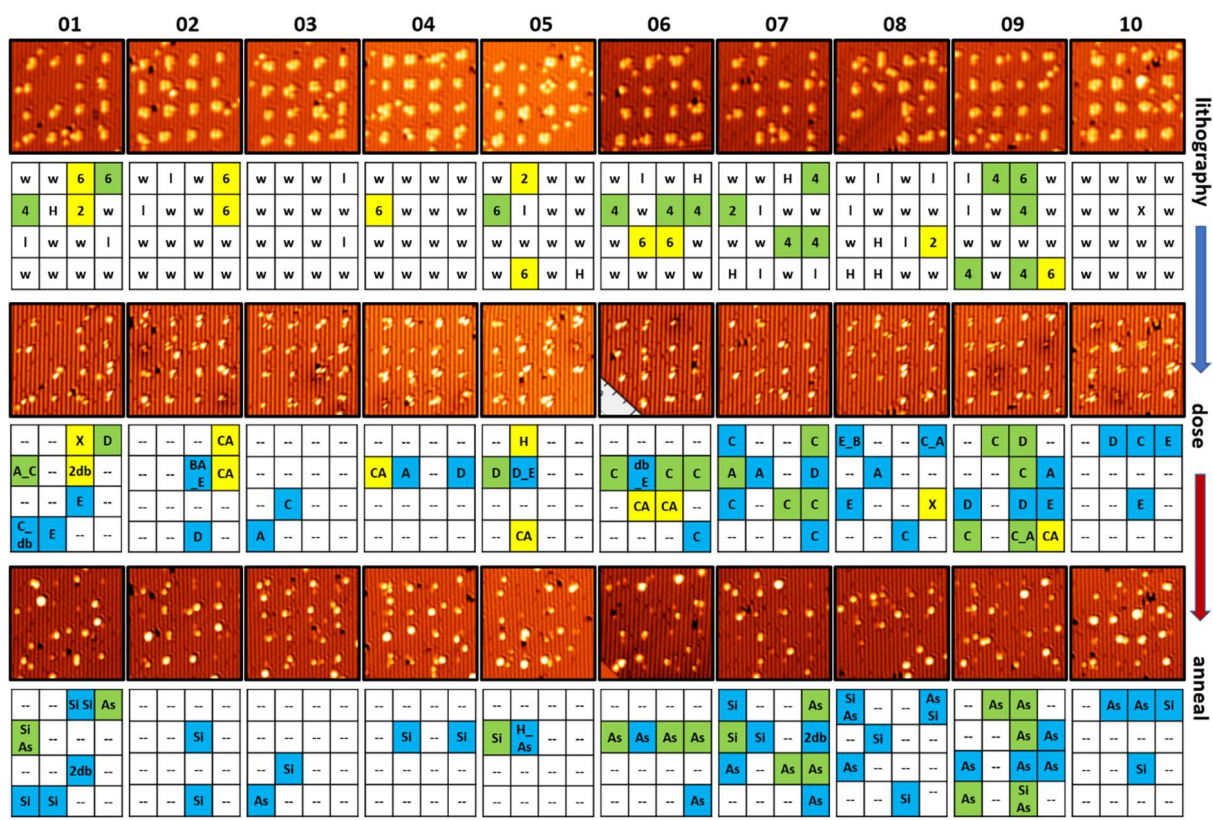

**Figure S4. Adsorption and Incorporation Statistics at Lithographic Adsorption Windows:** The 30 STM images above show ten 4×4 arrays patterned on a single atomic terrace of H:Si(001) shown in Figure S3. Each array is ~20×20 nm with a ~5 nm pitch. As indicated on the right-hand side of the image rows, all ten arrays (number at top of column) are followed through the three-step process of STM-HRL fabrication: lithographic patterning, arsine dosing, and thermal annealing for incorporation. At each process step, the array elements are assessed and results are tabulated in the grids immediately below each image; yellow grid cells are followed between process stages one and two, blue between stages two and three, and green are followed through all three stages. In the lithography stage, array elements are assessed for size and labelled "2", "4", or "6" indicating the number of DBs in a symmetric single-row lithographic window, or "l" or "w" indicating windows that are too long or too wide. In the dosing stage, all $AsH_x$ adsorbates, including clusters of adsorbates, are tabulated according to feature type "A", "C", "D", or "E", or "X" for contaminant. Two adsorbates in the same array element separated by at least one dimer row are indicated by an underscore. Finally, in the anneal stage, any isolated single adsorbate feature from the dosing images (including those that are found at a "w" or "l" window) are assessed for incorporation. The reaction products are recorded as "As", "Si", "H", or "db" indicating As-Si heterodimer, ad-Si, hydrogen termination, or dangling bond(s). Statistics in Figures 3i and 3l are generated using this procedure to evaluate the images above as well as data shown in Figures 3a and 3e.

23